# Quantifying lower-limb muscle coordination during cycling using electromyography-informed muscle synergies

**Reza Ahmadi**[1], Shahram Rasoulian[2], Hamidreza Heidary[1], Saied Jalal Aboodarda[2], Thomas K. Uchida[3], Walter Herzog[1,2], Amin Komeili[1,2]

[1]Department of Mechanical and Manufacturing Engineering, University of Calgary, Calgary, Canada
[2]Human Performance Laboratory, Faculty of Kinesiology, University of Calgary, Calgary, Canada
[3]Department of Mechanical Engineering, University of Ottawa, Ottawa, Canada
Email: Reza.Ahmadi3@ucalgary.ca

## Abstract

Assessment of muscle coordination during cycling may provide insight into motor control strategies and movement efficiency. This study evaluated muscle synergies and coactivation patterns as indicators of neuromuscular coordination in lower-limb across three power levels of cycling. Twenty recreational cyclists performed a graded cycling test on a stationary bicycle ergometer. Electromyography was recorded bilaterally from seven lower-limb muscles and muscle synergies were extracted using non-negative matrix factorization. The Coactivation Index (CI), Synergy Index (SI), and Synergy Coordination Index (SCI) were calculated to assess muscle coordination patterns. Four muscle synergies were identified consistently across power levels, with changes in synergy composition and activation timing correlated with increased muscular demands. As power level increased, the CI showed reduced muscle coactivation at the knee and greater muscle coactivation at the ankle. The SI revealed a greater contribution of the synergy weights of the extensor muscles than those of the flexor muscles at the knee. In contrast, the relative EMG contribution of hip extensor and flexor muscles remained consistent with increasing power levels. The SCI increased significantly with increasing power level, suggesting a reduction in the size of the synergy space and improved neuromuscular coordination. These findings provide insight into how the central nervous system modulates its response to increasing mechanical demands. Combining synergy and coactivation indices offers a promising approach to assess motor control, inform rehabilitation, and optimize performance in cycling tasks.

**Keywords: Biomechanics, Cycling, Muscle Coordination, Muscle Synergy, Electromyogram**

## 1. Introduction

Analysis of lower-limb joint motion is essential to understand human locomotion and activities such as cycling, to optimize human performance, and to develop injury prevention protocols [1]. Electromyography (EMG) signal parameters, such as the coactivation index (CI) and co-contraction index, have been used to quantify and compare muscle activity in agonist and antagonist muscle pairs [1], [2], [3], [4]. The CI measures the relative activation between opposing muscle groups (e.g., flexors and extensors) in a planar movement, while the co-contraction index assesses the degree of simultaneous activation of paired muscles, reflecting joint stability and stiffness. However, these metrics cannot fully represent the synergists involved in joint motions and have been used primarily to analyze muscles in a single limb [5].

Human movement is orchestrated by the neural dynamics of the central nervous system (CNS), proprioceptors, and muscles. The inherent redundancy in the human musculoskeletal system grants the CNS the flexibility to select the most appropriate movement and muscle coordination for a given task from a vast set of possibilities [6], [7]. This capability is crucial for generating adaptive behaviors in daily life. The CNS organizes numerous degrees of freedom and resolves actuator redundancy in the musculoskeletal system [7]; by coordinating muscle groups, the CNS simplifies the control of these degrees of freedom [8]. Despite progress in computational modeling, such as the development of models explaining biological control principles [9], [10], [11], understanding how the CNS achieves efficient movement remains a challenge. One proposed mechanism is behavior training, which reduces redundancy and narrows the focus of the CNS to a smaller set of optimized movement options. Alnajjar et al. [12] demonstrated that training enhances stability and muscle coordination, enabling the CNS to optimize movement patterns and reduce unnecessary actions. This mechanism has important implications for neurorehabilitation as it suggests that targeted training may streamline motor control strategies and improve recovery outcomes in individuals with motor impairments.

Muscle synergies are linear combinations of individual muscle activations that the CNS employs to efficiently produce and control movement. Muscle synergy theory simplifies the high-dimensional muscle activation space into a lower-dimensional synergy space. Each synergy is defined by a fixed set of weights representing the relative magnitudes of individual muscle activations [13]. A synergy's activation profile is a time-dependent signal that shows how the synergy is recruited and modulated over time [13]. The number of synergies used to reconstruct the original EMG signals defines the "synergy level" or the dimensionality of the model [13], [14], [15]. The CNS organizes muscle activity into synergies, groups of muscles activated together, to efficiently control phases of movement, such as joint flexion and extension [5], [16].

A muscle synergy analysis can identify motor actions that are common across various joint motions; however, validation of muscle synergies as indicators of lower-limb movement performance remains limited. This limitation arises from the complexity of neural and muscular interactions, the variability of synergies across individuals and tasks, and the challenges in obtaining consistent and high-quality EMG data [17]. Expanding research to include diverse conditions, populations, and longitudinal studies is essential to establish broader applicability of muscle synergy analysis [5]. Decomposition methods such as non-negative matrix factorization (NNMF) can be used to extract synergies from EMG signals [17], [18]. Previous research [19], [20], [21], [22] demonstrated the superiority of NNMF over other dimensionality reduction methods, such as Principal Component Analysis (PCA) and Independent Component Analysis (ICA), particularly for analyzing neuromuscular actions [19], [20], [21], [22]. One key advantage of NNMF lies in its restriction to non-negative weights, which aligns with the physiological reality that muscle activations cannot be negative. Compared to PCA, NNMF not only captures neuromuscular patterns more effectively [20], [22], [23] but also provides results that are more clinically interpretable, offering insight into muscle synergies and coordination [24], [25]. These attributes make NNMF a preferred method for understanding and modeling neuromuscular control. Accuracy of a muscle synergy scheme in representing a group of muscle functions depends on the EMG signal quality of dominant muscles and the number of synergy levels, and can be quantified using metrics such as the Variance Accounted For (VAF) [26].

Muscle synergy analysis has been used to provide a simplified representation of motor control for many complex human movements. Examples include sit-to-stand [27], [28], walking [29], running [30], [31], reaching [32], [33], and catching [34]. Recent studies have also explored exercises such as treadmill walking [35], [36] and cycling [37], [38], [39], [40], supporting the hypothesis that simplified motor control strategies are employed during these activities. For instance, NNMF-based muscle synergies have been used to study muscle fatigue during squatting [41] and to investigate the effects of anterior cruciate ligament (ACL) injuries on muscle recruitment patterns during walking [42]. In general, these studies have found consistent synergy patterns across participants and tasks, reflecting shared neural mechanisms for motor control, while also identifying task-specific adjustments in muscle contributions to meet unique movement demands.

Cycling, as a rhythmic and kinematically constrained activity, provides a controlled environment for studying muscle synergies. Previous studies have demonstrated that a few robust and shared muscle synergies can describe muscle coordination during pedaling across both trained and untrained cyclists, and among both healthy individuals and those with incomplete spinal cord injuries [43], [44]. The CNS adapts its control mechanisms through flexible synergy combinations in response to kinematic changes (e.g., posture) and kinetic changes (e.g., joint torque) [43], [45]. For instance, Turpin et al. [44] demonstrated that distinct muscle coordination patterns during standing cycling and seated cycling allow the CNS to optimize muscle activation under varying power levels. Similarly, De Marchis et al. [45] demonstrated that although inter-individual variability exists in cycling coordination strategies, common synergies are consistently used, reflecting the ability of the CNS to adapt modular control strategies to meet biomechanical demands. However, these studies identified a fixed set of muscle synergies for pedaling under steady conditions and did not investigate in detail the dependence of synergy patterns on power level.

The aim of this study was to investigate muscle synergies and coactivation patterns as indicators of neuromuscular coordination in lower-limb joints during cycling at varying power levels. By analyzing the consistency of synergies across power levels, we quantified the phase shifts in synergy activation profiles with respect to crank angle. Additionally, we employed the Synergy Coordination Index (SCI) to evaluate the size of the synergy space, providing insight into the muscle coordination strategies [12], [46]. By incorporating the Synergy Index (SI) with other muscle coactivation indices, we aimed to assess how the CNS optimizes synergy structures in response to varying muscle force demands. This study sought to extend the application of SCI and SI beyond static tasks to a dynamic and rhythmic cycling task, offering new perspectives on motor control adaptation and potential clinical implications for movement assessment and rehabilitation.

## 2. Methodology

The methodological framework was designed to systematically assess neuromuscular coordination during a cycling task with incrementally increasing power output. The experiments encompassed participant preparation, bilateral EMG recording from key lower-limb muscles, and a structured stepwise power protocol. As illustrated in Figure 1, the experimental workflow included EMG acquisition, signal processing, muscle synergy extraction, and clustering, culminating in the calculation of three synergy-based indices: SI, SCI, and CI. This comprehensive approach enabled the evaluation of how muscle coordination strategies evolve across a range of power levels during cycling. The limbs were analyzed separately to account for potential neuromuscular asymmetries.

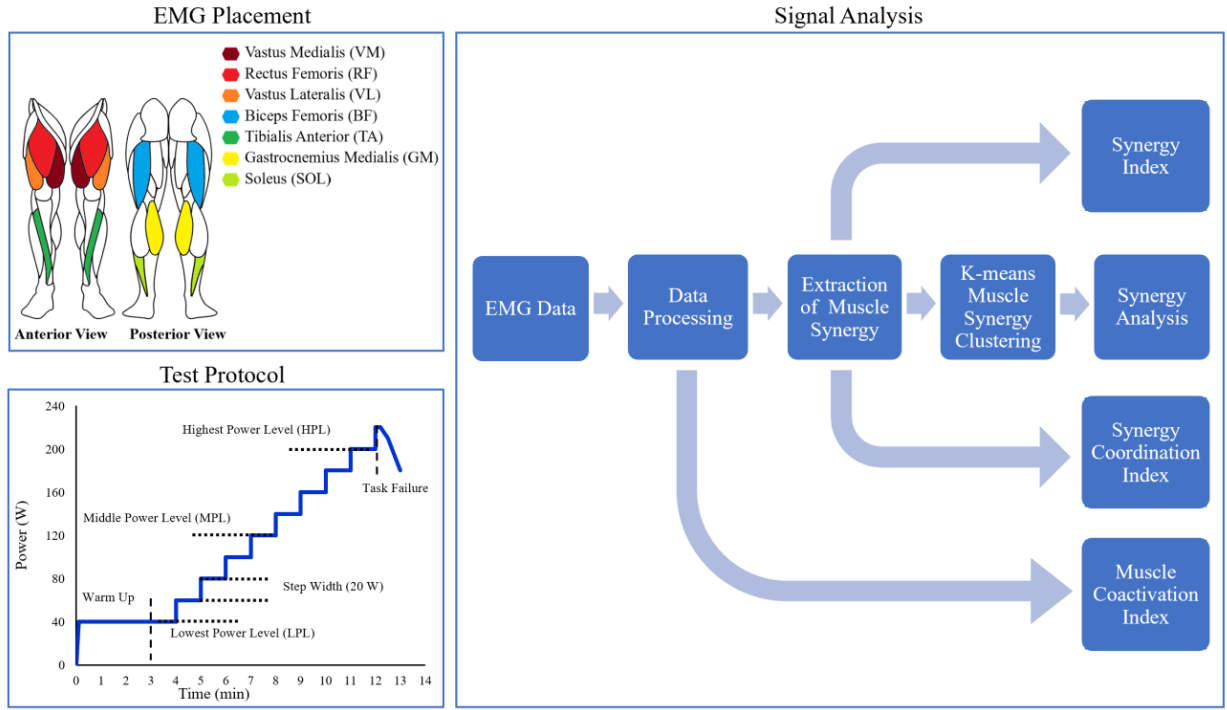

Figure 1. Overview of experiments for quantifying lower-limb movement neuromuscular control during cycling.

## 2.1. Participants

This study included 20 adult amateur cyclists, comprising 8 males and 12 females (age $22.6 \pm 4.5$ years, height $171.8 \pm 7.7$ cm, mass $70.6 \pm 12.5$ kg; mean $\pm$ standard deviation). The inclusion criterion was the absence of any diagnosed locomotor or cardiopulmonary pathology that could hinder participants' ability to complete the test or influence the results. All participants provided written informed consent prior to the study, which was conducted at the Human Performance Laboratory of the University of Calgary. Ethics approval was obtained from the University of Calgary Ethics Board (REB #23-1803).

## 2.2. Experimental protocol

Experiments were performed on a stationary cycling ergometer trainer (Echelon Connect EX-5, Echelon, Vancouver, Canada). A crank encoder (Model LM13, RLS, Komenda, Slovenia) measured crank angle at a sampling frequency of 250 Hz and was used to normalize the EMG signals to crank cycles. Participants were provided with cycling shoes with cleats; multiple sizes were made available for comfort. Cleats were aligned with the fat pad at the first metatarsal head and their orientation was validated using cardboard shoe insoles to ensure consistency between the two feet.

EMG signals were recorded bilaterally from seven lower-limb muscles: vastus medialis (VM), rectus femoris (RF), vastus lateralis (VL), biceps femoris long head (BF), tibialis anterior (TA), gastrocnemius medialis (GM), and soleus (SOL). For each muscle, the subscripts "D" and "ND" (e.g., $TA_D$) denote the dominant and non-dominant legs, respectively. EMG data were collected based on the European "SENIAM" recommendations [47]. Briefly, the sensor location on the skin was shaved, cleaned with alcohol pads, and allowed to dry before sensor placement. EMG signals were collected at a sampling rate of 2.2 kHz using a wireless Trigno EMG system (Delsys Inc., Natick, MA, USA), and recorded via Trigno Discover software (version 1.7.0). EMG data were synchronized with the encoder data to enable the analysis of muscle activation patterns relative to crank angles.

Participants completed a step test followed by a 3-minute warm-up period of cycling at 40 W and a cadence of 70 rpm. Following the warm-up period, the power level was increased in 20 W increments every minute, starting from 40 W, until the participant either failed to sustain the prescribed power level or decided to terminate the test. Power levels were displayed on a monitor and participants were verbally instructed to maintain the target power level.

## 2.3. Data Processing and Analysis

To facilitate comparison across participants, EMG data were normalized with respect to the dynamic maximum voluntary contraction (MVC). The MVC was calculated for each muscle as the average of the top 10% of EMG magnitudes recorded during the test [49]. To remove the effects of power level transitions, the first and last 10% of cycles at each power level were excluded from the analysis. The remaining 80% of the EMG data, comprising approximately 56 cycles per power level, were averaged over the crank angle to account for time shifts in EMG signals across cycles due to small variations in pedaling cadence. To monitor synergy variation across power levels, synergy parameters were calculated at three key points: the lowest power level (LPL), the middle power level (MPL), and the highest power level (HPL). The MPL was defined as half the maximum number of power levels (rounded down to the nearest integer if necessary).

The normalized EMG signals were high-pass filtered (30 Hz, zero-lag, fifth-order Butterworth), detrended, full-wave rectified, and low-pass filtered (10 Hz, zero-lag, fourth-order Butterworth) to obtain linear envelopes [38], which were then downsampled from 2.2 kHz to 250 Hz. The resulting filtered and downsampled EMG signals were processed to extract synergy weights, which were used to compute two indices: the SI and the SCI. The SI reflects the overall contribution of muscle synergies to joint motions; the SCI quantifies the size of the synergy space, capturing the adaptability and complexity of muscle coordination strategies. These two indices provide a comprehensive assessment of muscle synergy and coordination during cycling tasks.

## 2.4. Synergy Analysis

The filtered and downsampled EMG signal obtained from each muscle was first divided by its standard deviation to compute unit variance, ensuring equal weighting of the muscles in the synergy extraction process [48]. Muscle synergies were then extracted from the EMG data using NNMF with a multiplicative update rule:

$$\text{EMG}_{m \times n}^{\text{Exp}} = \begin{bmatrix} x_1^{VM_D} & \cdots & x_n^{VM_D} \\ \vdots & \ddots & \vdots \\ x_1^{SOL_{ND}} & \cdots & x_n^{SOL_{ND}} \end{bmatrix} \quad (1)$$

$$\text{EMG}_{m \times n}^{\text{Exp}} = \text{EMG}_{m \times n}^{\text{Rec}} + E_{m \times n}^{\text{EMG}} \quad (2)$$

$$\text{EMG}_{m \times n}^{\text{Rec}} = W_{(m \times k)} H_{(k \times n)} \quad (3)$$

where $\text{EMG}_{m \times n}^{\text{Exp}}$ contains the processed experimental EMG data recorded at each power level during pedaling (e.g., $X^{VM_D}$); $\text{EMG}_{m \times n}^{\text{Rec}}$ is the reconstructed version of $\text{EMG}_{m \times n}^{\text{Exp}}$, estimated using NNMF; $E_{m \times n}^{\text{EMG}}$ is the reconstruction error; $W_{(m \times k)}$ are the synergy weights; and $H_{(k \times n)}$ are the activation profiles. Here, m indicates the muscle ($VM_D$, $RF_D$, $VL_D$, $BF_D$, $TA_D$, $GM_D$, $SOL_D$, $VM_{ND}$, $RF_{ND}$, $VL_{ND}$, $BF_{ND}$, $TA_{ND}$, $GM_{ND}$, and $SOL_{ND}$), n denotes the EMG data point at each power level, and k represents the synergy level. Initial non-negative factors $W_0$ and $H_0$ were computed using the multiplicative update algorithm with 100 iterations and four synergy levels (k = 4), repeated ten times [49]. Additionally, the alternating least squares algorithm was employed with 1000 iterations and ten repetitions to enhance the accuracy of the resulting non-negative factors [5].

The criteria used to determine the number of muscle synergy levels included a combination of global and local variability of muscle activations, constrained by VAF, as suggested by Safavynia and Ting [50]:

$$\text{VAF} = 1 - \frac{\left\| E_{m \times n}^{\text{EMG}} \right\|_F^2}{\left\| \text{EMG}_{m \times n}^{\text{Exp}} \right\|_F^2} \quad (4)$$

$$\text{Criteria:} \begin{cases} \text{VAF}_{\text{Global}} \geq 0.95 \\ \text{VAF}_{\text{Local}} \geq 0.85 \quad \text{or} \quad R^2 \geq 0.6 \end{cases} \quad (5)$$

where $\|\ldots\|_F$ denotes the Frobenius norm of the matrix. The smallest number of synergies (k) that satisfied the VAF-based criteria was set as the decomposition level for the NNMF method. This approach ensured that the extracted synergies effectively captured the majority of the variance in muscle activation patterns while maintaining localized accuracy. After calculating the muscle synergy weights (W) and activation profiles (H), each muscle's contribution within a synergy, reflecting its relative activation level, was quantified using the corresponding weight values in W. To determine the reliability of these contributions, 95% confidence intervals were computed for each muscle's weight within each synergy using a bootstrapping procedure. Muscles were considered significant contributors to a synergy if their 95% confidence interval did not include zero. Relative activation strength was computed as the normalized weight of each muscle within a synergy and was used to rank muscles by their relative activation levels. These rankings were then compared across participants and synergies using cumulative magnitude analysis, providing insight into the dominant muscle groups involved in each synergy [51].

Because the obtained synergies appeared in an arbitrary order across participants, k-means clustering was used to label the synergies [52], [53]. To increase consistency within each cluster, participant data were excluded if two synergies were assigned to the same cluster or if the correlation of a synergy with the cluster mean was below 0.61, in accordance with $\alpha = 0.02$ for a sample size of 14 [54]. To ensure the k-means solution was globally optimal and independent of initial conditions, clustering was repeated 100 times using different randomly selected initial centroids and the result that was obtained most frequently was retained.

To investigate the temporal changes in synergy activation profiles across power levels, we employed a cross-correlation-based estimate of phase shift [55]. This method quantified phase shifts, in degrees, between synergy activation profiles that were common across power levels. To further examine the role of synergies in cycling mechanics, the revolution of the crank was divided into four quadrants: early downstroke (ED, 0° to 90°), late downstroke (LD, 90° to 180°), early upstroke (EU, 180° to 270°), and late upstroke (LU, 270° to 360°). Angles were measured with respect to the location of the pedal of the non-dominant leg, with 0° corresponding to the top dead center. This segmentation enabled a detailed analysis of muscle activation patterns throughout the pedaling cycle.

## 2.5. Indices for Analyzing Muscle Synergy

### 2.5.1. Synergy Index (SI)

To calculate SI, a synergy structure was formed by selecting the largest synergy weight for each of the fourteen muscles across the four synergy levels:

$$\begin{Bmatrix} VM_D \\ RF_D \\ VL_D \\ BF_D \\ \vdots \\ BF_{ND} \\ TA_{ND} \\ GM_{ND} \\ SOL_{ND} \end{Bmatrix} = \begin{Bmatrix} \max(W_{VM_D}^{k1}, W_{VM_D}^{k2}, W_{VM_D}^{k3}, W_{VM_D}^{k4}) \\ \max(W_{RF_D}^{k1}, W_{RF_D}^{k2}, W_{RF_D}^{k3}, W_{RF_D}^{k4}) \\ \max(W_{VL_D}^{k1}, W_{VL_D}^{k2}, W_{VL_D}^{k3}, W_{VL_D}^{k4}) \\ \max(W_{BF_D}^{k1}, W_{BF_D}^{k2}, W_{BF_D}^{k3}, W_{BF_D}^{k4}) \\ \vdots \\ \max(W_{BF_{ND}}^{k1}, W_{BF_{ND}}^{k2}, W_{BF_{ND}}^{k3}, W_{BF_{ND}}^{k4}) \\ \max(W_{TA_{ND}}^{k1}, W_{TA_{ND}}^{k2}, W_{TA_{ND}}^{k3}, W_{TA_{ND}}^{k4}) \\ \max(W_{GM_{ND}}^{k1}, W_{GM_{ND}}^{k2}, W_{GM_{ND}}^{k3}, W_{GM_{ND}}^{k4}) \\ \max(W_{SOL_{ND}}^{k1}, W_{SOL_{ND}}^{k2}, W_{SOL_{ND}}^{k3}, W_{SOL_{ND}}^{k4}) \end{Bmatrix} \quad (6)$$

where the subscripts and superscripts of the synergy weights indicate, respectively, the muscle name and the corresponding synergy group. The joint-specific SI was defined as the ratio of the sum of the largest synergy weights of the flexor muscles ($W_{flex}$) to the sum of the synergy weights of all muscles that generate moments at that joint:

$$SI = \frac{W_{flex}}{W_{flex} + W_{ext}} \quad (7)$$

The SI values for the hip ($SI_{hip}$), knee ($SI_{knee}$), and ankle ($SI_{ankle}$) joints were calculated as follows:

$$SI_{hip} = \frac{W_{RF}}{W_{RF} + W_{BF}} \quad (8)$$

$$SI_{knee} = \frac{W_{BF} + W_{GM}}{W_{BF} + W_{GM} + W_{VM} + W_{RF} + W_{VL}} \quad (9)$$

$$SI_{ankle} = \frac{W_{TA}}{W_{TA} + W_{GM} + W_{SOL}} \quad (10)$$

### 2.5.2. Synergy Coordination Index (SCI)

SCI was used to evaluate the dimensionality of the resulting synergy space, which quantifies the coordination among the extracted synergies. The synergy weight matrix W can be expressed as follows:

$$W = [W^{(1)} \ldots W^{(k)}] \quad (11)$$

where each $W^{(i)} \in \mathbb{R}^m$ is a basis vector of the synergy space, with k = 4. Since the NNMF algorithm was used to estimate $W^{(i)}$, the synergy vectors had only non-negative components. In general, the vectors $W^{(i)}$ are not mutually orthogonal. The size of the synergy space depends on the angles between the vectors that form its spanning set. To quantify the size of the synergy space, the Synergy Coordination Index (SCI) was defined using the average of the pairwise inner products between the normalized synergy vectors:

$$\text{SCI} = \frac{1}{P(k,2)} \sum_{i \neq j}^{k} (W^{(i)} \cdot W^{(j)}) \tag{12}$$

where $P(k,2)$ denotes the number of 2-permutations of k elements (with k = 4 in the present study), and "·" indicates the vector dot product. The SCI ranges from 0 to 1. An SCI of 1 indicates that the $W^{(i)}$ vectors are identical, meaning that the synergy space has collapsed into a line. Conversely, an SCI of 0 indicates that the $W^{(i)}$ vectors are mutually orthogonal and the synergy space is of maximal size. Thus, as the SCI increases, the synergy space becomes smaller due to greater cosine similarity between synergy vectors; a small value of the SCI indicates a larger synergy space and higher dimensionality in muscle coordination.

### 2.5.3 Coactivation Index (CI)

The CI of the hip, knee, and ankle joints ($CI_{hip}$, $CI_{knee}$, and $CI_{ankle}$, respectively) was computed as the ratio of the integrated EMG signals (iEMG) from the flexor muscles to the total integrated EMG signals from both flexor and extensor muscles:

$$CI_{hip} = \frac{iEMG_{RF}}{iEMG_{RF} + iEMG_{BF}} \tag{13}$$

$$CI_{knee} = \frac{iEMG_{BF} + iEMG_{GM}}{iEMG_{BF} + iEMG_{GM} + iEMG_{RF} + iEMG_{VM} + iEMG_{VL}} \tag{14}$$

$$CI_{ankle} = \frac{iEMG_{TA}}{iEMG_{TA} + iEMG_{GM} + iEMG_{SOL}} \tag{15}$$

where the EMG signals were preprocessed by filtering and downsampling, and then integrated over the crank angle [5], [16]. In this study, the focus was on critical phases in the pedaling cycle that were characterized by substantial mechanical demands at the hip, knee, and ankle joints. These critical phases were considered as the periods during which each joint experienced at least 80% of its maximum net moment. Based on joint moment data reported in prior studies [56], [57], high mechanical demand was observed at the hip between 70° and 170° of crank rotation, measured from the top dead center (TDC = 0°). Similar demands were reported for the knee from 15° to 75°, and for the ankle from 80° to 150°. A CI value greater than 0.5 indicates greater EMG activity from the flexor muscles while a value less than 0.5 indicates greater EMG activity from the extensor muscles [16], [53].

### 2.6. Statistical Analysis
The normality of the data was assessed using the Shapiro–Wilk test and the assumption of normality was confirmed (p > 0.05 for all variables). A one-way repeated-measures ANOVA was conducted to examine changes across three power levels [58]. For comparisons between two specific power levels, paired-sample t-tests were performed. Unless otherwise stated, any reported statistically significant difference refers to a comparison with the lowest power level. When significance refers to a comparison between other pairs of power levels, the specific groups are indicated. Statistical significance was set at p < 0.05. All analyses were conducted using Python (version 3.13).

## 3. Results

Figure 2 presents an example of processed EMG signals from one participant.

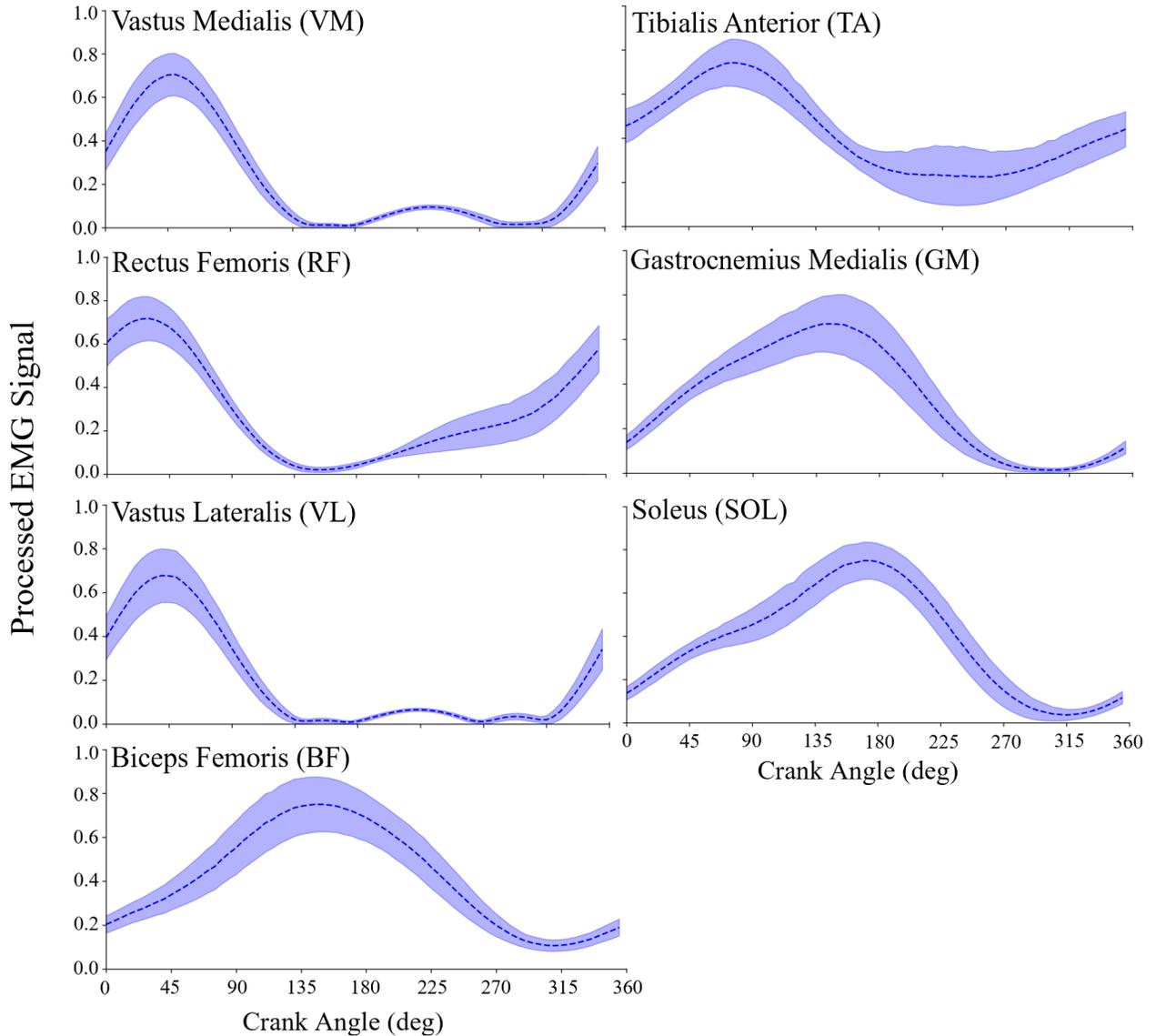

Figure 2. Processed EMG signals of non-dominant lower-limb muscles from one participant cycling at the 200 W power level (top dead center = 0°). Mean (dashed line) and standard deviation (shaded region) were computed over 56 cycles.

## 3.1. Muscle Synergies

The minimum number of synergies that satisfied the criteria of Eq. (5) was k = 4 (Figure 3). Applying the four identified synergy structures to the processed EMG data across all power levels resulted in a reconstruction accuracy of $VAF_{Global} > 0.95$ and $VAF_{Local} > 0.70$, $R^2 > 0.62$. These findings confirm that four muscle synergies provide a robust lower-dimensional representation of muscle coordination patterns across all power levels.

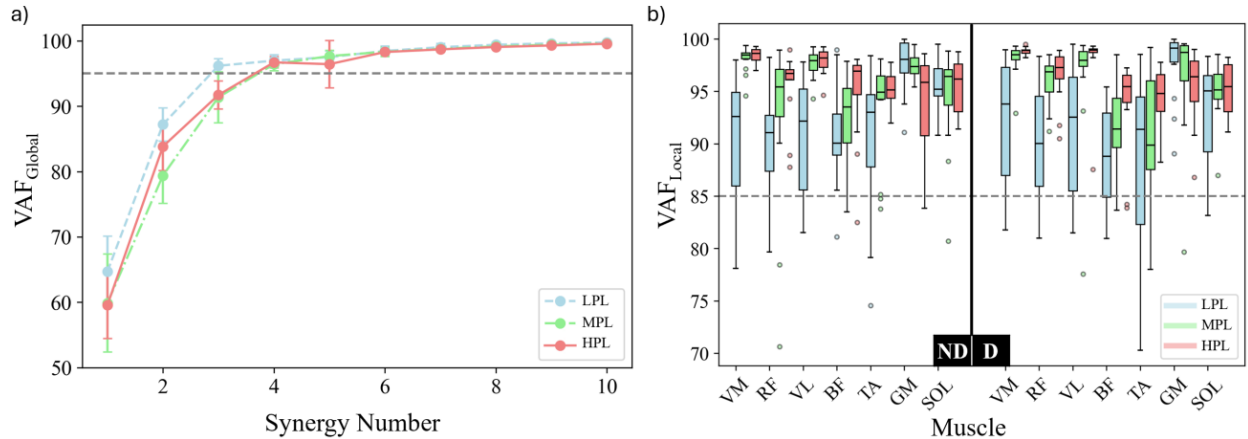

Figure 3. Muscle synergy selection based on Variance Accounted For (VAF) and $R^2$ criteria. **a)** $VAF_{Global}$ plotted against the number of muscle synergies for three power levels: lowest (LPL), middle (MPL), and highest (HPL). A minimum of four synergies (k = 4) was required to exceed the VAF threshold (gray dashed line) across all power levels. **b)** $VAF_{Local}$ across each muscle for the non-dominant (ND) and dominant (D) limbs. Each boxplot shows the distribution for LPL, MPL, and HPL conditions. The gray dashed line indicates the local VAF threshold; colored circles indicate outliers.

Synergy 1 (Figure 4a) primarily included the non-dominant quadriceps, which were active during the ED phase. Additionally, $SOL_D$ and $TA_{ND}$ were active during this phase. The muscle composition of Synergy 1 varied notably across power levels. In the LPL group, the synergy was dominated by $VL_{ND}$, $VM_{ND}$, $SOL_D$, $TA_{ND}$, and $RF_{ND}$. At MPL, $VL_{ND}$ and $VM_{ND}$ exhibited significantly greater activity than at LPL ($p < 0.05$). At HPL, $BF_D$ activity was significantly greater than it was at LPL ($p < 0.05$), indicating greater hamstring involvement at greater cycling intensity. Moreover, as the power level increased, the contribution of $SOL_D$ decreased significantly ($p < 0.05$).

Synergy 2 (Figure 4b) primarily included muscles that were active during the LD phase. The main contributors across all power levels were $GM_{ND}$, $TA_D$, and $TA_{ND}$, muscles known for stabilizing the ankle and facilitating smooth force transfer [59]. $SOL_{ND}$ and $BF_{ND}$ were also key contributors in both the LPL and MPL groups. At HPL, $BF_{ND}$ activity significantly increased compared to LPL ($p < 0.05$) and exhibited greater weight than $SOL_{ND}$; in contrast, $GM_{ND}$ weight significantly decreased from LPL to HPL. These changes indicate increased posterior thigh muscle activity during increased cycling intensity.

Synergy 3 (Figure 4c) primarily included muscles that were active during the EU phase. The primary contributors to Synergy 3 ($VM_D$, $VL_D$, $RF_D$, $SOL_{ND}$, and $SOL_D$) were consistent at LPL and MPL. However, at HPL, the weight of $RF_D$ increased ($p < 0.05$) while the contributions from $SOL_{ND}$ and $SOL_D$ decreased ($p < 0.05$), indicating a shift toward quadriceps activity during the EU phase. Synergy 4 (Figure 4d) was contralateral to Synergy 2 and consisted primarily of muscles that were active during the LU phase. The primary contributors to Synergy 4 ($GM_D$, $SOL_D$, $BF_D$, $TA_D$, and $TA_{ND}$) were consistent across all power levels, supporting pedal recovery and limb elevation. At HPL, the weights of $RF_{ND}$ and $TA_{ND}$ increased whereas the weight of $GM_D$ decreased ($p < 0.05$).

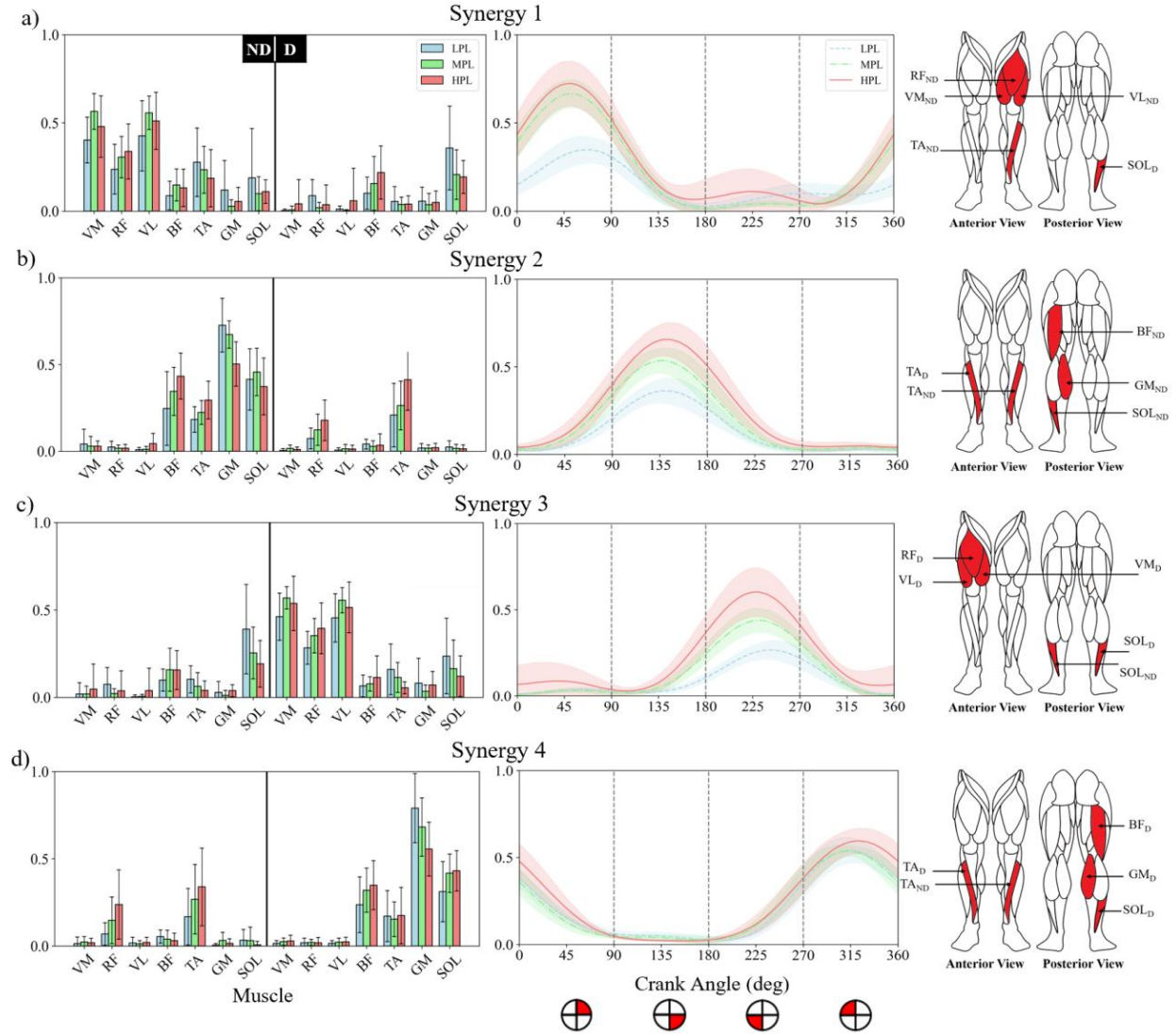

Figure 4. Weights (left) and activation profiles (center) of muscle synergies during cycling at three power levels for the non-dominant (ND) and dominant (D) legs. LPL: lowest power level (60 W); MPL: middle power level (midway between LPL and HPL); HPL: highest power level. The pedaling cycle is divided into four phases: early downstroke (0° to 90°), late downstroke (90° to 180°), early upstroke (180° to 270°), and late upstroke (270° to 360°). Angles are measured with respect to the location of the pedal of the non-dominant leg, with 0° corresponding to the top dead center. The five most prominent contributors to each synergy are highlighted (right).

The synergy activation profiles showed strong correlations between LPL and MPL, and between LPL and HPL, with $r > 0.86 \pm 0.13$ across module pairs. However, a comparison of these correlations revealed that LPL-to-HPL consistently exhibited lower correlations than LPL-to-MPL (i.e., correlation between activation profiles decreased as power level increased). For example, in Synergy 4, the LPL-to-MPL correlation was $0.98 \pm 0.02$ and the LPL-to-HPL correlation was 0.88 ± 0.09. Cross-correlation analysis was used to identify phase shifts between activation profiles of the same synergy across power levels (Table 1). While a phase shift was observed in all synergies, the shift was statistically significant for only Synergy 2 and Synergy 4. In Synergy 2, the shift increased from 0.92° ± 1.80° to 2.38° ± 5.22° across power levels; in Synergy 4, the shift increased from 4.85° ± 4.32° to 17.92° ± 11.75°.

Table 1. Cross-correlation shift and Pearson correlation coefficients of activation profiles between power levels (LPL-to-MPL and LPL-to-HPL) for each muscle synergy (mean ± standard deviation).

|  | Cross-Correlation Shift (°) | | Pearson Correlation Coefficient | |
| --- | --- | --- | --- | --- |
|  | LPL-to-MPL | LPL-to-HPL | LPL-to-MPL | LPL-to-HPL |
| Synergy 1 | 1.15 ± 2.34 | 2.00 ± 4.14 | 0.91 ± 0.08 | 0.86 ± 0.13 |
| Synergy 2 | 0.92 ± 1.80 | 2.38 ± 5.22 | 0.98 ± 0.01 | 0.95 ± 0.05 |
| Synergy 3 | 10.23 ± 5.17 | 14.07 ± 10.80 | 0.95 ± 0.04 | 0.91 ± 0.10 |
| Synergy 4 | 4.85 ± 4.32 | 17.92 ± 11.75 | 0.98 ± 0.02 | 0.88 ± 0.09 |

### 3.2. Synergy Index (SI)

The synergy outcomes at the hip, knee, and ankle joints were quantified using the SI calculated from Eqs. (7)–(10) (Figure 5). Recall that SI values greater than 0.5 indicate a predominance of the synergy weights of the flexor muscles, while values below 0.5 indicate a predominance of the synergy weights of the extensor muscles. Joint-specific contributions were assessed across the three power levels.

At the hip, mean SI values were close to 0.5 across all power levels, indicating an approximately equal contribution of the synergy weights of the flexor and extensor muscles at the hip. For the non-dominant hip, SI values were 0.44 ± 0.08 at LPL, 0.47 ± 0.17 at MPL, and 0.49 ± 0.17 at HPL; for the dominant hip, SI values were 0.50 ± 0.11 at LPL, 0.56 ± 0.15 at MPL, and 0.54 ± 0.15 at HPL. Statistically significant asymmetry was detected at MPL ($p = 0.020$), indicating a greater contribution of the synergy weights of flexor muscles relative to extensor muscles on the dominant leg at the middle power level.

At the knee, a clear trend was observed from predominance of the synergy weights of the flexor muscles at low power to predominance of the synergy weights of the extensor muscles at higher power. For the non-dominant knee, SI decreased significantly from 0.57 ± 0.09 at LPL to 0.47 ± 0.11 at MPL ($p < 0.05$), and further to 0.42 ± 0.07 at HPL ($p < 0.05$). The dominant knee followed a similar pattern, with SI decreasing from 0.47 ± 0.07 at LPL to 0.36 ± 0.10 at MPL ($p < 0.05$), and further to 0.34 ± 0.05 at HPL ($p < 0.05$). Significant side-to-side differences were observed at all three power levels, with consistently greater extensor dominance in the dominant knee ($p < 0.05$ at all power levels).

At the ankle, a predominance of synergy weights of the extensor muscles (SI < 0.5) was observed on both legs across all power levels. Interestingly, SI values increased as power level increased, opposite to the pattern observed at the knee. For the non-dominant ankle, SI increased significantly from 0.15 ± 0.05 at LPL to 0.18 ± 0.08 at MPL, and further to 0.31 ± 0.12 at HPL ($p < 0.05$). The dominant ankle showed a similar trend, with SI increasing significantly from 0.19 ± 0.09 at LPL to 0.23 ± 0.08 at MPL, and further to 0.27 ± 0.10 at HPL (p < 0.05). At the ankle, no significant differences in SI were observed between sides, across power levels.

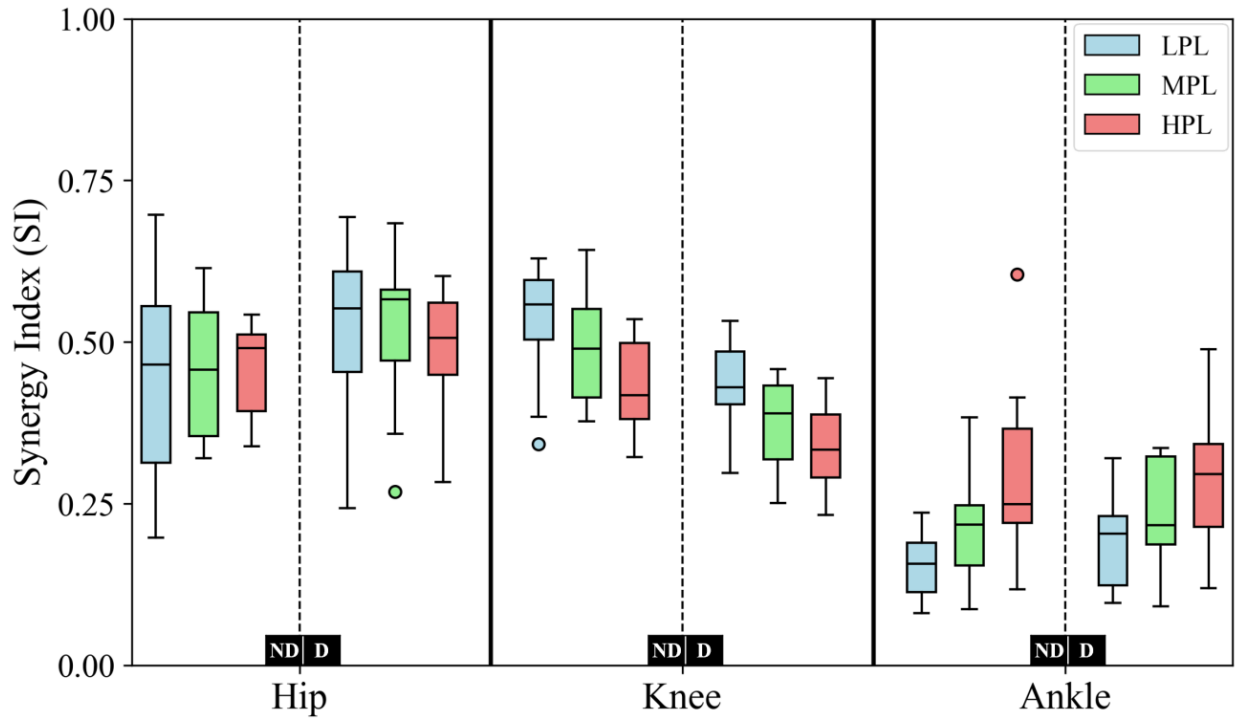

Figure 5. Synergy Index (SI) at the hip, knee, and ankle joints at three power levels for the non-dominant (ND) and dominant (D) legs. LPL: lowest power level (60 W); MPL: middle power level (midway between LPL and HPL); HPL: highest power level. Colored circles indicate outliers.

### 3.3. Synergy Coordination Index (SCI)

The SCI was calculated separately for each power level to assess changes in the structure of muscle coordination. SCI values showed a progressive increase with increasing power output: 0.08 ± 0.04 at LPL, 0.13 ± 0.08 at MPL, and 0.24 ± 0.11 at HPL. Statistical analysis confirmed significant differences among all three conditions (p < 0.05), demonstrating that power level had a significant effect on the size of the synergy space. This trend in the SCI indicates greater similarity among synergy vectors and a reduction in the dimensionality of muscle coordination at higher cycling intensities. In other words, as cycling intensity increased, muscle activation patterns became more consistent across the synergy space.

### 3.4. Coactivation Index (CI)

The CI for the hip, knee, and ankle is shown in Figure 6Figure 6 at all three power levels. For the hip, CI values were relatively consistent across power levels. On the non-dominant limb, CI values were 0.37 ± 0.20 (LPL), 0.30 ± 0.10 (MPL), and 0.30 ± 0.12 (HPL); on the dominant limb, CI values were 0.39 ± 0.18 (LPL), 0.37 ± 0.15 (MPL), and 0.39 ± 0.16 (HPL). Statistical analysis revealed no significant differences in hip joint synergy across power levels (p > 0.05). Between-limb differences in CI at the LPL were not significant at the hip. A significant difference appeared at MPL (p < 0.05) and became more pronounced at HPL (p < 0.02), indicating an increase in asymmetry with an increase in power level. In contrast, there were significant reductions in CI values at the knee as the power level increased. The non-dominant knee CI decreased from 0.33 ± 0.18 (LPL) to 0.22 ± 0.13 (MPL, $p < 0.05$) to 0.19 ± 0.09 (HPL, $p < 0.05$). Similarly, the dominant knee CI decreased from 0.28 ± 0.13 (LPL) to 0.18 ± 0.07 (MPL, $p < 0.05$) to 0.17 ± 0.06 (HPL, $p < 0.05$). These trends suggest increased recruitment of the knee extensor muscles at higher power levels. No significant between-limb differences in CI were found at the knee. At the ankle, CI values increased significantly as power level increased. On the non-dominant limb, the ankle CI increased from 0.19 ± 0.09 (LPL) to 0.23 ± 0.07 (MPL, $p < 0.05$) to 0.33 ± 0.11 (HPL, $p < 0.05$). On the dominant limb, the ankle CI increased from 0.13 ± 0.06 (LPL) to 0.19 ± 0.10 (HPL, $p < 0.05$), though the difference between MPL and HPL was not statistically significant. Significant between-limb differences in ankle CI were observed across all power levels ($p < 0.05$).

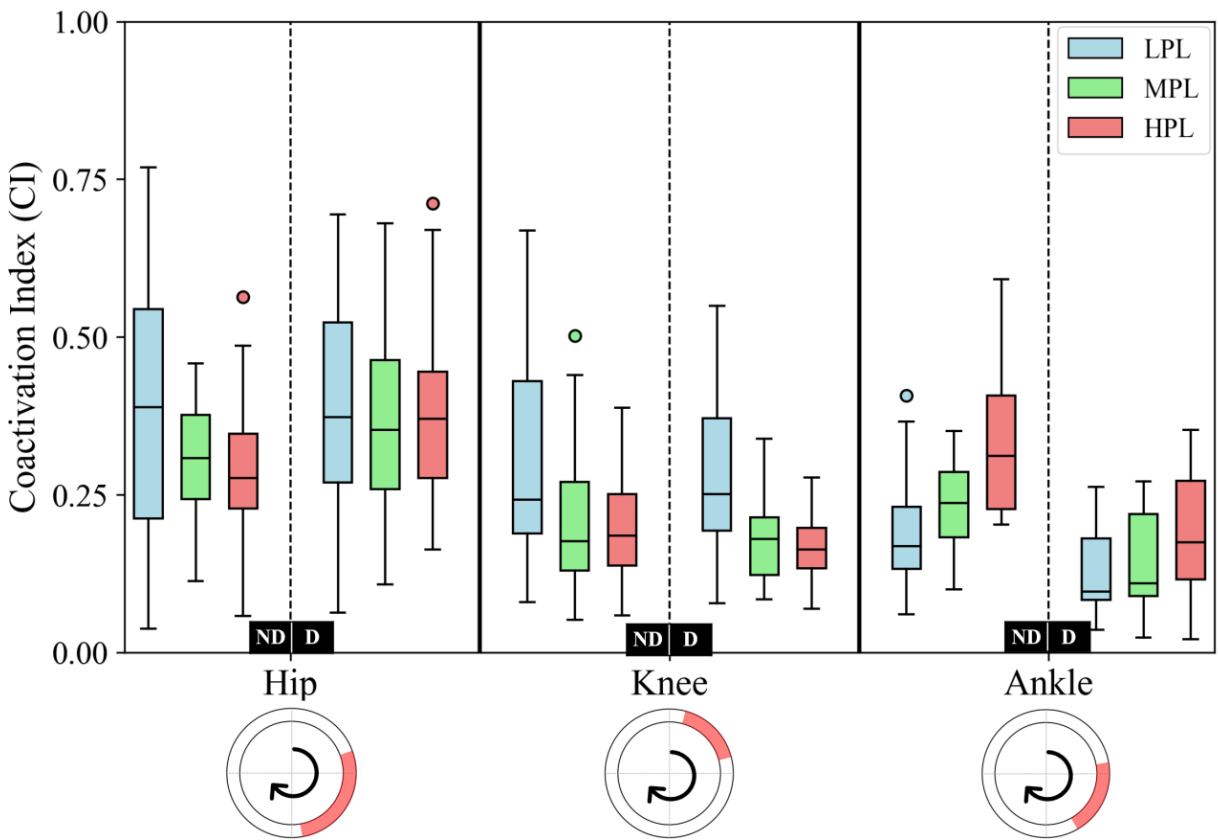

Figure 6. Coactivation index (CI) at the hip, knee, and ankle joints at three power levels for the non-dominant (ND) and dominant (D) limbs. LPL: lowest power level (60 W); MPL: middle power level (midway between LPL and HPL); HPL: highest power level. Colored circles indicate outliers. The circular schematics below each boxplot illustrate the crank cycle, with shaded arcs indicating periods during which each joint generated at least 80% of its peak net joint moment.

# 4. Discussion

## 4.1. Muscle Synergy Composition Across Power Levels

The results of this study indicate that muscle synergy structure during cycling depends on the power level. In terms of composition, similar synergies were observed across all power levels (LPL, MPL, and HPL), indicating preservation of the modular organization of muscle coordination. However, clear changes in muscle weighting within each synergy emerged with increasing intensity, as shown in Figure 4. For example, the first and third synergies, which involve primarily knee extensors and hip extensors, showed increased weighting of the hamstrings and quadriceps (particularly the rectus femoris) at higher power levels. This result suggests that there is greater reliance on these muscles to meet elevated torque demands at higher power levels. Conversely, the second and fourth synergies, which were dominated by the hamstrings and ankle dorsiflexors, demonstrated an increase in hamstring contribution as power level increased. This shift was accompanied by a significant reduction in gastrocnemius medialis weighting, suggesting a functional reorganization of the synergy structure to accommodate the demands of higher workloads.

Our hypothesis was that when pedaling at a constant cadence and the power demand is increased, propulsive muscle force must increase. Consequently, one might expect to observe phase shifts in muscle activations as it takes time to activate a muscle and allow its force to reach the required magnitude. Based on our results, significant workload-induced temporal adaptations were observed in the activation profiles of the second and fourth synergies across power levels, as evidenced by cross-correlation phase shifts and Pearson correlation coefficients between LPL, MPL, and HPL (Table 1). Synergy 2 exhibited moderate but statistically significant advanced phase shifts in activation timing with increased workload, showing a crank cycle shift of $2.38° \pm 5.22°$ between LPL and HPL ($p < 0.05$). Despite this phase shift, the Pearson correlation remained high, decreasing only slightly from $0.98 \pm 0.01$ (LPL to MPL) to $0.95 \pm 0.05$ (LPL to HPL), indicating relatively consistent activation patterns regardless of power level. In contrast, Synergy 4 showed more pronounced temporal adaptations, with statistically significant advanced phase shifts of $17.92° \pm 11.75°$ from LPL to HPL ($p < 0.05$). Correspondingly, the Pearson correlation decreased significantly from $0.98 \pm 0.02$ (LPL to MPL) to $0.88 \pm 0.09$ (LPL to HPL; $p < 0.05$), suggesting a notable reorganization of activation dynamics at higher power levels. Synergies 1 and 3 showed high Pearson correlation coefficients—Synergy 1: $0.91 \pm 0.08$ (LPL to MPL), $0.86 \pm 0.13$ (LPL to HPL); Synergy 3: $0.95 \pm 0.04$ (LPL to MPL), $0.91 \pm 0.10$ (LPL to HPL)—indicating a high degree of waveform similarity across workloads. However, no statistically significant phase shifts ($p > 0.05$) were observed for either Synergy 1 or Synergy 3 with increasing power level. These modest temporal adjustments agree with previous findings, which report only moderate timing shifts in primary power–generating synergies in response to increased cycling loads [37], [45]. This trend is expected, given the electromechanical delay between neural activation and force production [60], and may also vary across individuals due to differences in muscle properties, coordination strategies, and motor unit recruitment [45]. These changes in activation timing likely reflect the adaptive strategy used by the CNS to optimize muscle engagement and improve power transfer efficiency during high-intensity cycling. Such phase advancements and broader activation profiles have been well documented in the literature as characteristic responses to increasing mechanical demands during cycling [37], [61], [62].

The findings of our study agree with and extend the existing literature on muscle synergies during cycling. Hug et al. [43] found that muscle coordination during pedaling in trained cyclists could be explained by three synergies that were remarkably consistent across individuals. Notably, interindividual differences in surface EMG patterns did not correspond to differences in synergy structure; all cyclists employed similar synergies to generate pedaling force. This observation supports the hypothesis that muscle modules for cycling are task invariant. In our study, the similarity of synergies across cycling intensities further supports the task-invariant hypothesis of muscle modules. Indeed, synergy overlap across power levels was high: each LPL synergy could be matched to a counterpart at MPL and HPL, consistent with the high between-condition correlations observed in Table 1 for most synergies. Similarly, prior work by Esmaeili and Maleki [40] reported high similarity in muscle synergy vectors (average correlation of approximately 0.88) across a range of cadences and resistances. Together, these results suggest that the CNS relies on a stable set of muscle groupings during cycling, even as mechanical demands vary. However, our findings also reveal that within this stable set, the relative weighting of individual muscles may vary with power level. This observation aligns with previous studies showing that the required effort affects the relative contribution of individual muscles. For example, Enders et al. [63] reported that doubling the pedaling workload from 150 W to 300 W led to shifts in individual muscle contributions, while the overall activation patterns were largely preserved. In our data, the effect of increased workload on individual muscle contributions was observed as increased weighting of the quadriceps muscles, particularly the rectus femoris, and a concurrent decrease in gastrocnemius medialis contribution at HPL compared to LPL (Figure 4), indicating muscle-specific adaptations in synergy composition.

Overall, these findings demonstrate a flexible and nuanced temporal modulation by the CNS, selectively adjusting the activation profiles of certain synergies, particularly Synergies 2 and 4, to effectively meet the increased mechanical demands when power level was increased.

### 4.2. Flexor–Extensor Dominance and Functional Adaptation across Power Levels

In the context of this study, the SI quantifies the relative contribution of the synergy weights associated with flexor and extensor muscle groups at a joint, effectively indicating whether a joint's motor output is dominated by flexors or extensors. Liu et al. [64] demonstrated that walking speed influences the lower-limb SI: the index was greater at faster speeds than slower speeds, suggesting that the balance between flexor and extensor muscle activity depends on the mechanical demand. Similarly, in our study, SI values changed systematically with power level: as workload increased, the neuromuscular system adjusted the recruitment of flexors versus extensors to meet task demands. In essence, SI offers a robust window into how the CNS adjusts muscle group contributions to match the power requirements of the task.

Previous research has shown that during maximum-effort cycling, power production is distributed approximately as follows: 40% from the hip, 40% from the knee, 15% from the ankle, and 5% from the trunk and upper limbs [65]. This distribution highlights the hip's substantial contribution to propulsion. We observed a balanced relative contribution of the synergy weights associated with flexor and extensor muscle groups at the hip (i.e., SI of approximately 0.5) across power levels, attesting to the hip's central role in postural control and efficient force transmission. A significant asymmetry was observed at MPL (SI = 0.56 ± 0.15; Figure 5), which may reflect the fact that the dominant limb often exhibits greater capacity for neuromuscular control than the non-dominant limb (e.g., people are more coordinated when kicking a ball with their dominant leg than with their non-dominant leg) [66].

At the knee, a pronounced transition from greater synergy weights of the flexor muscles at low cycling intensity (SI > 0.5 at LPL) to greater synergy weights of the extensor muscles at higher intensities (SI < 0.5 at MPL and HPL) was observed, particularly on the dominant side (LPL: SI = 0.34 ± 0.05; Figure 5). As the power demand increased, muscle activity shifted toward the quadriceps (knee extensors), which are primarily responsible for producing torque during the downstroke. This pattern aligns with previous EMG studies that found disproportionately greater quadriceps activation at higher power levels [67], [68]. The persistent dominance of the dominant-knee extensors ($p < 0.05$ across all conditions) may reflect side-specific strength differences or pedaling mechanics that favor the dominant limb during power generation. Such asymmetry, especially in recreational cyclists, may increase the risk of overuse injuries, highlighting the importance of targeted bilateral training to restore balance [69]. These findings are consistent with motor control principles suggesting that the CNS prioritizes the activation of the main power-generating muscles during high-demand activities (e.g., quadriceps during HPL) while also minimizing antagonist coactivation to maximize net joint torque and metabolic efficiency.

The ankle's consistently extensor-dominant Synergy Index (SI < 0.5) across all power levels underscores the critical role of the ankle plantarflexors (gastrocnemius medialis and soleus) in generating propulsive force during cycling. However, the significant increase in SI as cycling intensity increased (Figure 5) indicates increased recruitment of dorsiflexors (primarily tibialis anterior) to stabilize the ankle during forceful plantarflexion. This coactivation may enhance joint stiffness and control during high-load conditions. The absence of significant side-to-side differences ($p > 0.05$) points to a symmetrical control strategy at the ankle, possibly due to its more constrained and mechanically consistent role in the kinetic chain compared to the hip and knee [70]. This adaptable pattern, prioritizing extensor-driven propulsion while integrating flexors for stabilization, reflects the hierarchical approach of the CNS to balance force output with joint integrity.

### 4.3. Neuromuscular Control Strategy Adaptations at Higher Power

The SCI offers valuable insight into understanding neuromuscular control strategy adaptations. Alnajjar et al. [12] introduced SCI as a measure of how similar the synergy activations are, with larger SCI values indicating a smaller synergy space (i.e., a greater overlap or similarity among synergies). In this context, a greater SCI reflects a more focused neural control strategy, characterized by the consistent recruitment of specific combinations of muscles. Training and skill acquisition have been associated with increases in SCI and related metrics, suggesting that experts exhibit more stable synergy composition and timing than novices [12]. In our study, increasing power level produced a similar effect as increasing skill demand: it led to greater coordination and reduced variability among synergies, and earlier, more synchronized synergy profiles, culminating in a greater SCI. This pattern suggests that exposure to high-intensity cycling may itself serve as a form of training, promoting a more expert-like neuromuscular coordination profile. Conversely, failure to adapt synergy coordination at higher intensities could signal a less flexible or less developed control strategy. Overall, the neuromuscular adaptations observed in our study reflect a hierarchical control model in which the CNS maintains a core set of muscle synergies, likely generated by central pattern generators, and modulates their activation timing and weighting in a context-sensitive manner. This combination of preserved modular structure and flexible modulation exhibits the hallmarks of efficient motor control: balancing the stability of learned coordination patterns with the adaptability required to meet changing mechanical demands.

## 4.4 Joint-specific Coactivation Patterns and Neuromuscular Strategies during Cycling

The CI at the hip remained similar across power levels. This observation suggests that agonist–antagonist muscle engagement at the hip is regulated in a consistent manner to maintain pelvic stability and enable efficient force transfer over a range of workloads. However, a significant difference in CI between the non-dominant and dominant hips at HPL may point to subtle asymmetries that emerge only under high mechanical demand. One possible explanation is limb dominance (i.e., a habitual preference for one leg), which may become more pronounced under intense conditions. Alternatively, these differences may reflect slight variations in technique, muscle fatigue, or compensation strategies adopted when sustaining high power levels. Minor side-to-side differences in muscle coordination are not uncommon in cycling. Although well-trained cyclists generally exhibit more symmetric pedaling force profiles [71] and more consistent synergy patterns than novices [43], subtle bilateral differences in muscle activation and coordination variability have been reported [61], [72].

In contrast to the hip, a clear decrease in CI was observed at the knee as cycling intensity increased. Significantly lower CI values were observed at the knee at HPL compared to LPL, indicating more efficient neuromuscular control under high mechanical demand. This observation suggests that as power demand increases, the CNS reduces antagonist muscle activity (e.g., hamstring engagement that would otherwise oppose quadriceps contraction during the critical phase from 15° to 75°), thereby optimizing the production of net joint torque. (However, it is important to note that biarticular muscles, such as the hamstrings, may still need to be active to generate extension moments at the hip joint, and their contribution depends on both crank position and inter-joint coordination.) This interpretation aligns with the broader motor control literature, which associates high-efficiency or skilled movement with low coactivation. For example, Chapman et al. [73] observed that less-trained cyclists exhibited greater coactivation at higher cadences whereas highly trained cyclists maintained low coactivation regardless of cadence. Similarly, the observed reduction in knee CI from LPL to HPL in our study suggests that participants adopted a more power-efficient activation strategy under greater workloads, selectively reducing unnecessary antagonist activation to improve movement economy.

At the ankle, the CI increased as cycling intensity increased, which was opposite to the trend observed at the knee. At low intensity, the muscles crossing the ankle exhibited relatively low coactivation, but as workload increased, so too did the antagonistic activity of the ankle dorsiflexors. Increasing CI at the ankle as cycling intensity increased may reflect a neuromuscular strategy aimed at increasing joint stability under greater force demands. Co-contraction increases joint stiffness, which is beneficial at the ankle during forceful pedaling because a stiffer ankle more effectively transmits power and absorbs greater forces from the pedal [74], [75]. In essence, the body "braces" the ankle by activating both agonists and antagonists to increase joint stiffness. Use of antagonistic coactivation to improve joint stability during high-demand tasks is well documented in the biomechanics literature [74]. For example, simulations of gait on slippery surfaces have shown that increased lower-limb coactivation enhances stability [76]. In general, muscle co-contraction is known to increase joint stiffness which results in reduced joint excursions upon a perturbation, but this strategy comes at the expense of an increased energetic cost [74], [75], [76]. The side-to-side differences in ankle CI, where the dominant ankle consistently showing greater extensor activation than the non-dominant, may reflect limb-specific neuromuscular strategies. While leg dominance is one potential explanation, other factors such as differences in neuromuscular control, joint stiffness regulation, or habitual movement asymmetries may also contribute. Although our participants were not trained athletes, prior studies have shown that even trained individuals can exhibit leg-specific neuromuscular patterns [72]. Together, these findings suggest that the CNS tailors coactivation not only to task intensity but also potentially to individual limb-specific control strategies, ensuring that ankle stability is prioritized under high power demands and revealing that the muscles on one limb may coactivate more to achieve that stability.

This study has several limitations. First, it was conducted on a relatively small and homogeneous sample of young, healthy recreational cyclists, which may limit the generalizability of the findings. While the chosen participant group reduced inter-participant variability, the observed muscle coordination strategies may not extend to other populations, such as trained athletes, older adults, or individuals with neuromuscular impairments. Second, all experiments were performed in a controlled laboratory setting using a stationary cycle ergometer, which does not replicate the physiological and environmental complexity of real-world cycling. Factors such as terrain variability, postural adjustments, fatigue accumulation, balance demands, and wind, which were absent in our experimental setup, may influence muscle coordination patterns. Future studies should aim to evaluate synergy coordination and joint-level indices in more natural conditions, including prolonged and outdoor cycling tasks. Third, surface EMG was used to record activity from a small number of superficial muscles, excluding deeper muscles as well as some key contributors to pedaling mechanics, such as the gluteus maximus. EMG is also susceptible to signal crosstalk and provides only a qualitative estimate of the corresponding muscle forces, making synergy-based interpretations only approximations of the underlying neuromuscular propulsive force and torque output. Finally, the joint-specific indices (CI, SI, and SCI) were derived from EMG signals without concurrent measurements of joint kinematics or kinetics. As such, we were unable to directly assess how muscle activity and coordination correlate with joint torques, pedal forces, or mechanical efficiency. Future research should integrate EMG with motion capture, force sensors, or musculoskeletal modeling to provide deeper insight into the biomechanics of cycling and to validate these indices as indicators of motor control efficacy in both healthy and clinical populations.

## 5. Conclusion

This study investigated muscle coordination strategies during cycling by analyzing muscle synergies and joint-specific coactivation patterns at three power levels, providing insights into how the CNS modulates neuromuscular control in response to increasing mechanical demands. The results demonstrate that while the core synergy structure remained consistent across intensities, underscoring the stability of modular motor control, distinct changes in muscle weighting, activation timing, and muscle coordination emerged, particularly within synergies associated with the late downstroke and late upstroke phases.

At the joint level, unique coactivation and flexor–extensor dominance patterns were observed. The hip maintained similar coactivation and balanced synergy contributions across all power levels, providing evidence of its role in postural support and force transmission. At the knee, reduced coactivation and a clear transition toward extensor dominance were observed at higher intensities, suggesting a shift toward more efficient torque generation. In contrast, at the ankle, increased coactivation was observed with increasing workload, perhaps serving to stabilize the joint under greater loads. These findings highlight joint-specific neuromuscular adaptations that collectively support whole-body performance during cycling. The SI further revealed task-specific changes in flexor–extensor dominance: consistent balance between flexor and extensor activity at the hip, extensor dominance at the knee, and extensor dominance at the ankle. Importantly, an observed increase in the SCI with increasing power level suggests that higher-intensity cycling evokes a smaller and more coordinated synergy space, indicative of enhanced neuromuscular efficiency and a possible shift toward expert-like control strategies.

Together, the CI, SI, and SCI offer complementary perspectives on how the nervous system regulates motor complexity, coordination, and joint-specific muscle contributions during dynamic tasks. This study advances our understanding of CNS organization under varying task demands and highlights the potential of synergy- and coactivation-based indices as indicators of neuromuscular function. These findings may inform applications in performance training, injury prevention, and rehabilitation by supporting individualized interventions aimed at improving motor efficiency and joint stability across a wide range of populations.